\documentstyle{ichep}
\newcommand\sss{\scriptscriptstyle}

\newcommand\muf{\mu_{\sss F}}
\newcommand\mur{\mu_{\sss R}}

\renewcommand\pt{\mbox{$p_{\sss \rm T}$}}
\newcommand\as{\alpha_{\sss S}}

\def	\be		{\begin{equation}}
\def	\ee		{\end{equation}}
\def	\ba		{\begin{eqnarray}}
\def	\ea		{\end{eqnarray}}

\def	\=		{\;=\;}
\def	\frac		#1#2{{#1 \over #2}}

\def	\to		{\rightarrow }

\def	\b		{\mbox{$b$}}

\def	\jpsi		{\mbox{$\psi$}}
\def	\psip		{\mbox{$\psi^\prime$}}

\def	\pt		{\mbox{$p_T$}}

\def	\as		{\mbox{$\alpha_s$}}

\begin{document}

\title{Recent Progress in the Theory of \\
Heavy Quark and Quarkonium Production\\
in Hadronic Collisions \\
{\large{hep-ph/9410299}}   }

\author{Michelangelo L Mangano$^{\S\|}$}

\affil{
\S\ Istituto Nazionale di Fisica Nucleare, Pisa ITALY}

\abstract{We review heavy quark and quarkonium production in high energy
hadronic collisions. We  discuss the status of the theoretical calculations and
their uncertainties. We then compare the current theoretical results
with the most recent measurements from the Tevatron Collider
experiments}

\twocolumn[\maketitle]

\fnm{4}{Address after Febr.~1, 1995: CERN, TH Division, Geneva, Switzerland.}

Heavy quark production in high energy hadronic collisions consitutes a
benchmark process for the study of perturbative QCD. The comparison of
experimental data with the predictions of QCD provides a necessary check that
the ingredients entering the evaluation of hadronic processes (partonic
distribution functions and higher order corrections) are under control and can
be used to evaluate the rates for more exotic phenomena or to extrapolate the
calculations to even higher energies. Likewise, production of quarkonium
states, in addition to provide yet another interesting framework for the study
of the boundary between perturbative and non-perturbative QCD,
is important in view of the possible use of exclusive charmonium decays of $b$
hadrons for the detection of CP violation phenomena.

In this presentation we review the current status of theoretical
calculations, and discuss the implications of the most recent experimental
measurements of $b$ quarks and charmonium states
performed at the Tevatron $p\bar p$ Collider.
For more complete reviews, including a discussion of heavy quark production at
fixed target energies, see Refs.~\cite{newfmnr,bdfm}.

\section{Open Flavour Production: Theory Overview}
To start with, we briefly report on the current status of the theoretical
calculations. One has to distinguish between calculations performed at a
complete but fixed order in perturbation theory (PT), and those performed
resumming classes of potentially large logarithmic contributions which arise at
any order in PT. The exact matrix elements squared for heavy quark production
in hadronic collisions are fully known up to the ${\cal O}(\as^3)$, both for
real and virtual processes. These matrix elements have been used to evaluate at
NLO the total production cross section \cite{btot}, single particle inclusive
distributions \cite{bpt}\ and two particle inclusive distributions (a.k.a.
correlations) \cite{mnr}.

Three classes of large logarithms can appear in the perturbative expansion for
heavy quark production:
\begin{enumerate}
\item $[\as\log(S/m_Q^2)]^n \sim [\as\log(1/x_{Bj})]^n$ terms, where $S$ is the
hadronic CM energy squared. These small $x$ effects are possibly relevant for
the production of charm or bottom quarks at the current energies, while should
have no effect on the determination of the $top$ cross section, given the large
$t$ mass. Several theoretical studies have been performed \cite{smallx}, and
the indications are that \b\ production cross sections should not increase by
more than 30-50\% at Tevatron energies due to these effects.
\item $[\as\log(m_Q/p^T_{QQ})]^n$ terms, where $p^T_{QQ}$ is the transverse
momentum of the heavy quark pair. These contributions come from the multiple
emission of initial state soft gluons, similarly to standard Drell Yan
corrections. These corrections have been studied in detail in the case of top
production, where the effect is potentially large due to the heavy top mass
\cite{laenen}. They are not relevant for the redefinition of the total cross
section of \b\ quarks, but affect the kinematical distributions of pairs
produced just above threshold \cite{berger2}, or in regions at the edge of
phase space, such as $\Delta\phi=\pi$.
\item $[\as\log(p_T/m_Q)]^n$ terms, where $p_T$ is the transverse momentum of
the heavy quark. These terms arise from multiple collinear gluons emitted by
a heavy quark produced at large transverse momentum, or from almost collinear
branching of gluons into heavy quark pairs. Again these corrections are
not expected to affect the total production rates, but will contribute to the
large \pt\ distributions of c and b quarks. No effect is expected for the top
at current energies. These logarithms can be resummed using a fragmentation
function formalism. A first step in this direction was taken by Cacciari and
Greco \cite{greco}, who convoluted the NLO fragmentation functions for heavy
quarks \cite{NasonMele}\ with the NLO parton level cross
section for production of massless partons\cite{nlojet}.
A significant improvement in the stability w.r.t. scale changes has been
observed for $\pt>50$ GeV.
\end{enumerate}

%
\section{Single Inclusive Bottom Production}
The status of $b$ production at hadron colliders has been quite puzzling for
some time. Data collected by UA1\cite{ua1b} at the CERN Collider ($\sqrt S$=630
GeV) were in good agreement with theoretical expectations based on the NLO QCD
calculations\cite{bpt}.
On the contrary, the first measurements performed at 1.8
TeV by the CDF\cite{cdfpre93} experiment at the Fermilab Collider showed a
significant discrepancy with the same calculation.

Owing to recent progress, the situation has considerably clarified.
The latest results from the Fermilab 1.8 TeV $p\bar
p$ Collider have been presented at this Conference by CDF\cite{cdfdpf} and by
the new experiment, D0\cite{d0dpf}.
The current situation is summarized in Figs.~\ref{BottomAtCdf}
and~\ref{BottomAtD0}, showing a comparison of the theoretical expectations
with the results from CDF and D0 for integrated \pt\ distributions of $b$
quarks.

The theoretical curves require some explanation. First of all, they do not
differ much from the original prediction\cite{bpt} using the DFLM
structure functions. New structure function fits, including
the first results from HERA, have recently become available. We use in our
prediction one of these sets, namely MRSA\cite{MRSA}. Since the values  of
$x$ probed by $b$ production at the Tevatron in the currently measured \pt\
range only cover the region $x> 5 \times 10^{-3}$, we observe no significant
change relative to the results obtained using older fits.

The second important point is the choice of a range for $\Lambda_5$.
Deep inelastic scattering results tend to favour small values of
$\Lambda$. For example, the set MRSA uses $\Lambda_5=151$~MeV.
On the other hand, LEP data favour a higher value: the
central value of $\Lambda_5$ at LEP is around 300 MeV. This value
is also supported by other lower-energy results, such as the $\tau$
hadronic width (for a review of $\Lambda_5$ determinations, see
the work by Catani\cite{CataniMarseille93}). It is therefore sensible to use
the range from 151 to 300 MeV for $\Lambda_5$.

The upper curves in Fig.\ref{BottomAtCdf}\ and Fig.\ref{BottomAtD0}\
correspond to the PDF set MRSA\cite{MRSA}, $\Lambda_5=300$ MeV, $m_b=4.5$ GeV
and $\mur=\muf=\sqrt{\pt^2+m_b^2}/2$. The lower curves correspond to
$\Lambda_5=151$ MeV, $m_b=5$ GeV and $\mur=\muf=2\sqrt{\pt^2+m_b^2}$.
In the absence of fits with $\Lambda_5$ frozen to the
desired values we chose to simply change the value of $\Lambda_5$ in the
partonic cross section. A discussion of this choice can be found in
\cite{newfmnr}.

Studies shown in Ref.~\cite{newfmnr}\ also indicate that
pre-HERA and post-HERA PDF sets predict $b$ cross sections which do not differ
by more than 5\% within a large range of \pt. While such a stability is
partly artificial, being related to the large overlap of correlated
measurements entering the determination of the parton distribution fits, it
however suggests that by now the uncertainty in the structure functions does
not leave much room by itself for significant changes in the expected $b$ cross
section at Tevatron energies.

Coming back to the comparison of theory and data, from Fig.\ref{BottomAtCdf}
we see that the CDF data points are
now consistent with the fixed-order theoretical prediction, although on the
high
side. The D0 points, instead, comfortably sit within the theoretical range.
In order to better compare data among themselves and with theory, we plot the
ratio between data points and the upper theoretical predition on a linear
scale (Fig.\ref{ratio}). From this figure we see that the UA1 and D0
data are well consistent with the upper theoretical curve, while CDF points are
slightly above. Until the apparent difference between D0 and CDF will be
understood, it is therefore appropriate to conclude that at present no
significant discrepancy between theory and data or between data at different
energies is being observed.
Once the experimental statistics and systematics will be further reduced,
it will be reasonable to assume that residual discrepancies of the same order
as those currently observed may be explained in terms of small-$x$ effects.
Additional theoretical studies of these effects, such as a better
understanding of the matching with the fixed-order
next-to-leading-order calculations, should therefore be pursued.




\section{Charmonium Production}
The $J/\psi$ and $\psi'$ states are of particular interest since
they are produced in abundance and are relatively easy to detect at a
collider such as the Tevatron.  In earlier calculations of direct charmonium
production at large transverse momentum ($p_T$) in $p \bar{p}$ collisions
\cite{onia}\ , it was assumed that the leading-order diagrams give the dominant
contributions to the cross section.  These calculations did not reproduce all
aspects of the available data \cite{ua1_psi,cdfpre93}, suggesting that there
are  other important production mechanisms.  It was pointed out by Braaten and
Yuan \cite{by1} in 1993 that fragmentation processes, while formally of  higher
order in the strong coupling constant $\alpha_{s}$, will dominate at
sufficiently large $p_T$.
The relevant fragmentation
functions for the production of the S-wave and P-wave states have all been
calculated to leading order in $\alpha_s$ (\cite{by1}-\cite{sf}).
Explicit calculations of the contribution to  $\psi$
production at the Tevatron from the fragmentation of gluons and charm  quarks
have recently been completed by several groups \cite{bdfm,cg,roy}

In Fig.\ref{PsiAtCdf}, the sum of the fragmentation
and of the leading-order contributions are
compared with  preliminary CDF data for prompt $\psi$ production
\cite{troy,cdfdpf}. \jpsi's from $\chi$ production and decay are included, both
in the theory curves and in the data.
The contribution to \jpsi\ production from $b$-hadron decays has instead been
removed from the data via detection of the secondary vertex from which the
\jpsi's originate \cite{troy,cdfdpf}.
While the shapes of the leading-order curve
and the fragmentation curve are both consistent with the data over the range
of $p_T$ that is available, the normalization of the leading-order
contribution is too small by more than an order of magnitude.  The
fragmentation contribution has the correct normalization to within a factor
of 2 or 3, which can be easily accounted for by the uncertainties of such a LO
calculation (for a discussion of these uncertainties, see \cite{bdfm}).
We conclude that the fragmentation calculation is not inconsistent
with the CDF data on prompt $\psi$ production.
A similar conclusion can be reached \cite{bdfm}, after inclusion of the
$b\to\psi$ contributions, from a comparison with the D0 data \cite{d0dpf}.

We
next consider the production of $\psi'$, which should not receive any
contributions from known higher charmonium states.  The $\psi'$ fragmentation
contribution can be obtained from the $g \to \psi$, $c \to \psi$, and
$\gamma \to \psi$ fragmentation contribution simply by multiplying by the
ratio of the electronic widths of the $\psi'$ and $\psi$.  The results are
shown in Fig.\ref{PspAtCdf}, along with
the  preliminary CDF data \cite{troy}.  Again the contribution from $b$--hadron
decays has been subtracted using the secondary vertex information.
In  striking contrast
to the case of $\psi$ production, the normalization of  the fragmentation
contribution to $\psi'$ production is too small by more  than an order of
magnitude.  That there is such a large discrepancy between  theory and
experiment in the case of $\psi'$, but not for $\psi$, is  extremely
interesting.  It suggests that there are other important mechanisms  for
production of S-wave states at large $p_T$ beyond those that have  presently
been calculated.  While such processes would certainly affect  \jpsi\
production as well, their effect may not be as dramatic because of  the large
contribution from $\chi_c$-production in the case of the \jpsi.

One possible such mechanism is the process $g g \to \psi g g$, with a gluon
exchanged in the $t$-channel, which we expect to be at least as large as the
direct and fragmentation processes calculated so far in the relevant region of
\pt. However, this would not be enough to explain the factor of 30 or so
observed discrepancy.
A more likely possibility is that as yet undetected higher charmonium states,
with significant BR's into \psip,
can be produced with large rates in $p \bar p$ collisions. At this meeting, an
interesting possibility was raised by F. Close \cite{close}: possible hybrid
charmonium states ($c\bar c g$ hadrons) are expected to have masses around 4.2
GeV, below the threshold for their only open charm decay to $\bar DD^{**}$. If
such states existed, they would have large BR's into $\psip\gamma$ or
$\psip\eta$. Other suggestions have also been made, including the possibility
of
2$^3P$ ($\chi$-like) states \cite{close,mwise,roy3}. Since the production rate
for these states would be very big, even relatively small BR's could easily
accomodate the current puzzling rate \cite{roy3}.
Searches for resonances in such channels are therefore encouraged.

\section{Conclusions}
Significant progress has taken place in this field over the past few years,
both in theory and experiments. The latest measurements at 1.8 TeV indicate an
acceptable agreement between the data and NLO QCD predictions for the \b\
inclusive \pt\ spectrum, and the presence now of two experiments will hopefully
reduce experimental uncertanties.
Previously detected discrepancies, observed in the inclusive \jpsi\ final
states, are now attributed to large sources of \jpsi\ direct production.
New theoretical work has explained the abundance of 1$^3S$ production (mostly
understood as coming from gluon fragmentation into $\chi$ states), but cannot
as yet account for the observed 2$^3S$ rate. The possibility that new and
possibly exotic charmonium states are being produced and observed at the
highest energies available today opens perhaps new interesting frontiers for
the already rich field of hadronic collider physics.

\section*{Acknowledgements} The work presented in this talk was carried out
in collaboration with E.~Braaten, M.~Doncheski, S.~Fleming, S.~Frixione,
P.~Nason and G.~Ridolfi. This work is supported in part by the
EEC Programme ``Human Capital and Mobility'', Network ``Physics at High Energy
Colliders'', contract CHRX-CT93-0357 (DG 12 COMA).

\Bibliography{99}
\def	\nuke	#1#2#3{{\em Nucl. Phys.} {\bf B#1}  (#2), #3}
\def	\pl  	#1#2#3{{\em Phys. Lett.} {\bf #1B}  (#2), #3}
\def	\prl  	#1#2#3{{\em Phys. Rev. Lett.} {\bf #1}  (#2), #3}
\def	\pr  	#1#2#3{{\em Phys. Rev.} {\bf #1}  (#2), #3}
\def	\prd  	#1#2#3{{\em Phys. Rev.} {\bf D#1}  (#2), #3}
\def	\zeit	#1#2#3{{\em Z. Phys.} {\bf C#1}  (#2), #3}
\def	\cmp 	#1#2#3{{\em Comm. Math. Phys.} {\bf #1}  (#2), #3}
\bibitem{newfmnr}
	S. Frixione, M. Mangano, P. Nason and G. Ridolfi,
	CERN-TH.7292/94 (1994), to appear in Nucl. Phys. B.
\bibitem{bdfm}
	E. Braaten, M. Doncheski, S. Fleming and M.L. Mangano,
	\pl{333}{1994}{548}.
\bibitem{btot}
	P.~Nason, S.~Dawson and R.~K.~Ellis,
	\nuke{303}{1988}{607};
	W.~Beenakker, H. Kuijf, W.L. van Neerven and J. Smith,
        \prd{40}{1989}{54}.
\bibitem{bpt}
	P.~Nason, S.~Dawson and R.~K.~Ellis,
	\nuke{327}{1988}{49 };
	W.~Beenakker et al.,
        \nuke{351}{1991}{507}.
\bibitem{mnr}
	M. Mangano, P. Nason and G. Ridolfi,
	\nuke{373}{1992}{295}.
\bibitem{smallx}
	J.C. Collins and R.K. Ellis,
	\nuke{360}{1991}{3};
	S. Catani, M. Ciafaloni and F. Hautmann,
	\nuke{366}{1991}{135};
	E.M. Levin, M.G. Ryskin and Yu.M. Shabelsky,
	\pl{260}{1991}{429}.
\bibitem{laenen}
	E. Laenen, J. Smith and W.L. van Neerven,
	\nuke{369}{1992}{543}.
\bibitem{berger2}
	E. Berger and R. Meng,
	\prd{49}{1994}{3248}.
\bibitem{greco}
	M. Cacciari and M. Greco, Univ. of Pavia FNT/T-93/43,
	hep-ph/9311260.
\bibitem{NasonMele}
	B. Mele and P. Nason,
	\nuke{361}{1991}{626}
\bibitem{nlojet}
	F. Aversa et al., \pl{210}{1988}{225}; \zeit{49}{459}{1991};\\
       	S. Ellis, Z. Kunszt, D. Soper, \prl{62}{1989}{2188}; {\bf 64} (1990),
	2121.
\bibitem{ua1b}
	C. Albajar et al., UA1 Coll., \pl{256}{1991}{121}.
\bibitem{cdfpre93}
  F. Abe et al., CDF Coll.,
  \prl{68}{1992}{3403}; 
  {\bf 69}(1992)3704; 
  {\bf 71}(1993)500,  
  2396  
  and 2537. 
\bibitem{cdfdpf}
	K. Byrum, CDF Coll.,
	presented at this Conference.
\bibitem{d0dpf}
     	D. Hedin and L. Markosky, D0 Coll.,
	presented at this Conference.
\bibitem{MRSA}
   A.D.~Martin, W.J.~Stirling and R.G.~Roberts,
   Rutherford Lab preprint RAL-94-055, DTP/94/34 (1994).
\bibitem{CataniMarseille93}
   S.~Catani, preprint DFF~194/11/93, to appear in the
   {\it Proceedings of the EPS conference},
   Marseille, 1993.
\bibitem{onia}
	E.L. Berger and D. Jones,
	\prd{23}{1981}{1521};\\
	R. Baier and R. R\"uckl, \zeit{19}{1983}{251};\\
	B. Humpert, \pl{184}{1987}{105};\\
	R. Gastmans, W. Troost and T.T. Wu, \nuke{291}{1987}{731};\\
	E.W.N. Glover, A.D. Martin and W.J. Stirling,
	\zeit{38}{1988}{473};\\
	E.W.N. Glover, F. Halzen and A.D. Martin,
	\pl{185}{1987}{441}.
\bibitem{ua1_psi}
 	C. Albajar et al., UA1 Coll., \pl{256}{1991}{112}.
\bibitem{by1} E. Braaten and T.C. Yuan,
                {\it Phys. Rev. Lett.} {\bf 71} (1993) 1673.
\bibitem{cc} C.-H. Chang and Y.-Q. Chen,
                {\it Phys. Lett.} {\bf B284} (1991) 127;
                {\it Phys. Rev.} {\bf D46} (1992) 3845.
\bibitem{wise} A.F. Falk, M. Luke, M.J. Savage, and M.B. Wise,
                {\it Phys. Lett.} {\bf B312} (1993) 486.
\bibitem{bcy} E. Braaten, K. Cheung, T.C. Yuan,
                {\it Phys. Rev.} {\bf D48} (1993) 4230.
\bibitem{by2} E. Braaten and T.C. Yuan,
                Fermilab preprint FERMILAB-PUB-94/040-T  (1994).
\bibitem{y} T.C. Yuan, U.C. Davis preprint UCD-94-2 (1994);
\bibitem{sf} 	S. Fleming,
                Fermilab preprint FERMILAB-PUB-94/074-T (1994).
\bibitem{cg} M. Cacciari and M. Greco,
                INFN preprint, FNT/T-94/13, hep-ph/9405241.
\bibitem{roy}
	D.P. Roy and K. Sridhar, CERN-TH.7329/94, hep-ph/9406386.
\bibitem{troy}
	CDF Coll., Fermilab-Conf-94/136-E.
\bibitem{close} F. Close, Rutherford Preprint, RAL 94-093 (1994).
\bibitem{mwise} P. Cho, M.B. Wise and S.P. Trivedi, FNAL-PUB 94/256-T.
\bibitem{roy3} D.P. Roy and K. Sridhar, CERN-TH.7434/94.
\end{thebibliography}
\end{document}